\documentclass[useAMS,usenatbib,usegraphicx]{mn2e}
\newcommand{\beq}{\begin{equation}}
\newcommand{\eeq}{\end{equation}}
\def\bea{\begin{eqnarray}}
\def\eea{\end{eqnarray}}

\def\gsim{~\rlap{$>$}{\lower 1.0ex\hbox{$\sim$}}}

\def\simpropto{\lower.2ex\hbox{$\; \buildrel \propto \over \sim \;$}}
\def\ltsim{\lower.5ex\hbox{$\; \buildrel < \over \sim \;$}}
\def\gtsim{\lower.5ex\hbox{$\; \buildrel > \over \sim \;$}}
\def\ltsim{\lower.5ex\hbox{$\; \buildrel < \over \sim \;$}}
\def\gtsim{\lower.5ex\hbox{$\; \buildrel > \over \sim \;$}}

\def\pmb#1{\setbox0=\hbox{#1}%
\kern-.025em\copy0\kern-\wd0
\kern.05em\copy0\kern-\wd0
\kern-.025em\raise.0433em\box0}

\def\simlt{\lower.5ex\hbox{$\; \buildrel < \over \sim \;$}}
\def\simgt{\lower.5ex\hbox{$\; \buildrel > \over \sim \;$}}

\def\fixit#1{}
\def\mpcc{{\rm {Mpc}^3}}

\usepackage{graphicx}
\usepackage{enumitem}
\usepackage{amssymb}
\usepackage{amsmath}
\usepackage{multirow}
\usepackage{mathrsfs}
\usepackage{bbding}
\usepackage{soul}
\usepackage{float}
\usepackage{color}
\usepackage[colorlinks,citecolor=cyan]{hyperref}
\usepackage{lineno}

\setstcolor{red}

\title[Explaining galaxy alignment over a range of scales]
      {A mechanism to explain galaxy alignment over a range of scales }
\author[Prabhakar Tiwari and Pankaj Jain]
 {Prabhakar Tiwari $^{1}$\thanks{ptiwari@nao.cas.cn},  
  Pankaj Jain$^{2}$\thanks{pkjain@iitk.ac.in} \\
$^{1}$ National Astronomical Observatories, CAS, Beijing 100012, China \\
$^{2}$ Department of Physics, Indian Institute of Technology, Kanpur-208016, India\\
        }
\begin{document}
\maketitle
\begin{abstract}
The observed large-scale alignment of polarization angles and galaxy axis have been challenging the fundamental assumption of homogeneity and isotropy in standard cosmology since more than two decades. The intergalactic magnetic field, and its correlations in real space, potentially seems as a viable candidate for explaining this phenomenon. It has been shown earlier that the large-scale intergalactic magnetic field correlations can explain the alignment signal of quasars over Gpc scale, interestingly they can also explain the radio polarization alignment observed in JVAS/CLASS data over 100 Mpc. Motivated with recent observations of galaxy axis alignment over several tens of Mpc, and Mpc scale, i.e., the cluster scale, we further explore the correlations of background magnetic field to explain these relatively small scale alignment observations.  In particular, we explore two recently claimed signals of alignment in the radio sources in the FIRST catalog and in the ACO clusters. We find that the FIRST alignment signal is well explained in terms of the intergalactic magnetic field with a spectral index of $-2.62\pm 0.03$. Furthermore, the model also partially explains the very small scale alignment (alignment within clusters). Though the elementary model proposed in this work seems to have its limitations at very small scales, the large-scale magnetic field correlations potentially seem to explain the polarization and galaxy axis alignment from Gpc to Mpc scales.
\end{abstract}

\section{Introduction}
\label{sec:intro}
There currently exists considerable evidence
that the galaxies  show alignment with one another over Mpc to Gpc distance scales, although 
the observations still lack firm confirmation. The first evidence was presented in \cite{Hutsemekers:1998} who found that the polarization of visible radiation from distant quasars is aligned over very large distance scales of order Gpc. The signal was confirmed by further investigations \citep{Hutsemekers:2000fv,Jain:2003sg,Hutsemekers:2005iz}. A signal of alignment of
radio polarizations has also been observed on smaller scales of order 100 Mpc \citep{Tiwari:2013pol,Pelgrims:2015}.  \cite{Mandarakas:2021} claim to find four regions of radio galaxy axis alignment on distance scale of order $\approx 400$ Mpc at a significance level of more than $5\sigma$. Furthermore several studies have claimed alignment of radio galaxy axis on distance scales of order 20 to 60 Mpc \citep{Taylor:2016rsd,Contigiani:2017,Panwar:2020}. However, the signal  of alignment appears to be absent in Astrogeo database radio jets \citep{Blinov:2020} and  in polarization of radio sources observed with IRAM (Institute for Radio Astronomy in the Millimeter Range) 30m telescope \citep{Tiwari:2018}.

There exist many theoretical attempts to explain these observations. It has been proposed that the effect is related to the correlation of magnetic field on cosmological \citep{Agarwal:2009ic,Agarwal:2012} or spercluster distance scales \citep{Tiwari:2016sp}. A hypothetical light pseudoscalar field has also been invoked to explain the large scale optical alignment \citep{Jain:2002vx,Agarwal:2009ic,Piotrovich:2009zz,Agarwal:2012}. One potential problem with this explanation is the relatively large circular polarization predicted by this phenomenon \citep{Hutsemekers:2010,Payez:2011}. However, there could be sources of decoherence, for example a fluctuating background  magnetic field can significantly reduced the circular polarization in comparison to linear polarization \citep{Agarwal:2012}. Other explanations include,  vector perturbations \citep{Morales:2007rd}, dark energy
\citep{Urban:2009sw}, cosmic strings \citep{Poltis:2010yu,Hackmann:2010ir}
violation of isotropy \citep{Ciarcelluti:2012pc}
and a superhorizon perturbation \citep{Chakrabarty:2016}.

Recently, \cite{Tovmassian:2020} explored the distribution of position angles (PA) of galaxies inside clusters and reported alignment, i.e. the galaxies' PAs are not oriented randomly. If the Universe is homogeneous and isotropic on large scales, we expect random orientations of galaxies if a large enough volume is considered. However, for cluster scales the local physical processes will dominate, and the  orientation of galaxies can be anisotropic. The alignment of galaxies within cluster is  presumably related to galaxy formation, the tidal gravitational fields in the large-scale structure can align nearby galaxies (\citealt{Tempel:2013,Zhang:2013, Kiessling:2015} and references therein). The orbital angular momentum and hence the background magnetic field must also play a role in such an alignment, which is so far not completely understood.
Furthermore, the visible galaxy shapes are distorted from weak-lensing and contribute to galaxy alignment \citep{Miralda-Escude:1991,Waerbeke:2000, Bartelmann:2001,Catelan:2001}. To explore galaxy formation there have been  many efforts over several decades for the study of the distribution of the PA of galaxies in clusters \citep{Sastry:1968,Adams:1980,Fong:1990,Fuller:1999,Chen:2019}.  \cite{Tovmassian:2020} further explore the galaxy alignment within clusters and assume 
alignment if, in a cluster, the galaxy number at one $90^\circ$ position angle interval is more than twice higher than the other $90^\circ$ interval and report relatively large alignment for galaxies in isolated clusters. \cite{Tovmassian:2020} propose that the galaxies in clusters could have been aligned at the beginning but with time the accretion and mutual interactions of galaxies could have randomized galaxy orientations, and thus in the rich clusters the alignment signal may not be significant.

The signal of alignment in radio galaxy PAs has also been probed \citep{Panwar:2020} in the Faint Images of the Radio Sky at Twenty-Centimeters (FIRST; \citealt{Becker:1995,2015ApJ...801...26H}) catalog. The authors report a significant signal of alignment for scales less than $1^\circ$ which corresponds to a distance scale of about 28 Mpc, consistent with what has been observed earlier \citep{Taylor:2016rsd}. With their 3D analysis, \cite{Panwar:2020} report mild alignment at larger distances.

Our basic idea is that there exists a background magnetic field which has spatial correlations over wide range of
length scales. This field interacts with the charged particles in the radio galaxies and tends to orient their jet axis along the background field direction. Hence the galaxy axes acquire the spatial correlations of the background field. 
Detailed numerical simulations are required in order to determine the precise nature of the induced correlations. Such simulations are so far not available in literature. However, we expect that a sufficiently strong background magnetic field acting for large enough time will have a tendency to align the individual galaxies within a cluster. Hence, as a simple approximation, we assume that the jet axes are aligned in exactly the same direction as the background magnetic field at the position of the galaxy. A more detailed study may relax this assumption by proposing a suitable distribution function of jet axes, which may lead to only partial correlations to the background field direction. We postpone this to future work.
We point out that this mechanism would also lead to correlations in the 
integrated radio polarizations. This is because these polarization angles are known to be correlated with the jet
axis \citep{Gabuzda:1994,Lister:2000,Pollack:2003,Helmboldt:2007,Joshi:2007}. The model is described in more detail in \cite{Tiwari:2016sp}. 

The origin of the background magnetic is assumed to be primordial and has correlations over wide range of length scales, including very large distances of
 order Gpc. Such correlations may explain the alignment signal of large distance scale quasar optical polarizations \cite{Hutsemekers:1998} by inducing
 intrinsic alignments in quasars on such large  distance scale. We assume that the magnetic field can be defined by a single power law model over these wide range of scales from Mpc to Gpc. Here we shall be interested in the application of the  model on alignment at relatively small distance scales of order less than 30 Mpc. 

Let $\vec B_i(\vec r)$ represents the $i^{\rm th}$ component of cluster magnetic field at position $\vec r$. We decompose it in terms of its Fourier components $b_i(\vec k)$, such that,
\begin{equation}
    B_i(\vec r) = \frac{1}{V} \sum_{\vec k} b_i(\vec k)e^{i\vec k\cdot \vec r}
\end{equation}
where $V$ is the volume.
It is reasonable to assume that the magnetic field is uncorrelated in Fourier space. Hence we can write the two point correlations as \citep{Tiwari:2016sp},
\begin{equation}
    \langle b_i^*(\vec k) b_j(\vec q) \rangle = \delta_{\vec k,\vec q} P_{ij}(\vec k) M(k)
\end{equation}
where $M(k)$ is the spectral function,
\begin{equation}
    P_{ij} = \left(\delta_{ij} - \frac{k_ik_j}{k^2}\right)
\end{equation}
and $k=|\vec k|$. The two point correlations in real space can be expressed as,
\begin{equation}
\label{eq:bcorr_real}
    \langle B_i(\vec r_0+\vec r) B_j(\vec r_0)\rangle = \frac{1}{V} \int d^3 k e^{i\vec k\cdot \vec r}P_{ij}(\vec k) M(k) W^2(kr_c) 
\end{equation}
where $W(kr_c)$ is a window function, 
\begin{equation}
W(x) = \left\{
    \begin{array}{l l}
      1  & x<1 \\
      0  & x>1
    \end{array}\right.
\end{equation}
and $r_c=k_c^{-1}$ is a suitably chosen cutoff scale.

\begin{figure}
    \centering
    \includegraphics[width=\columnwidth]{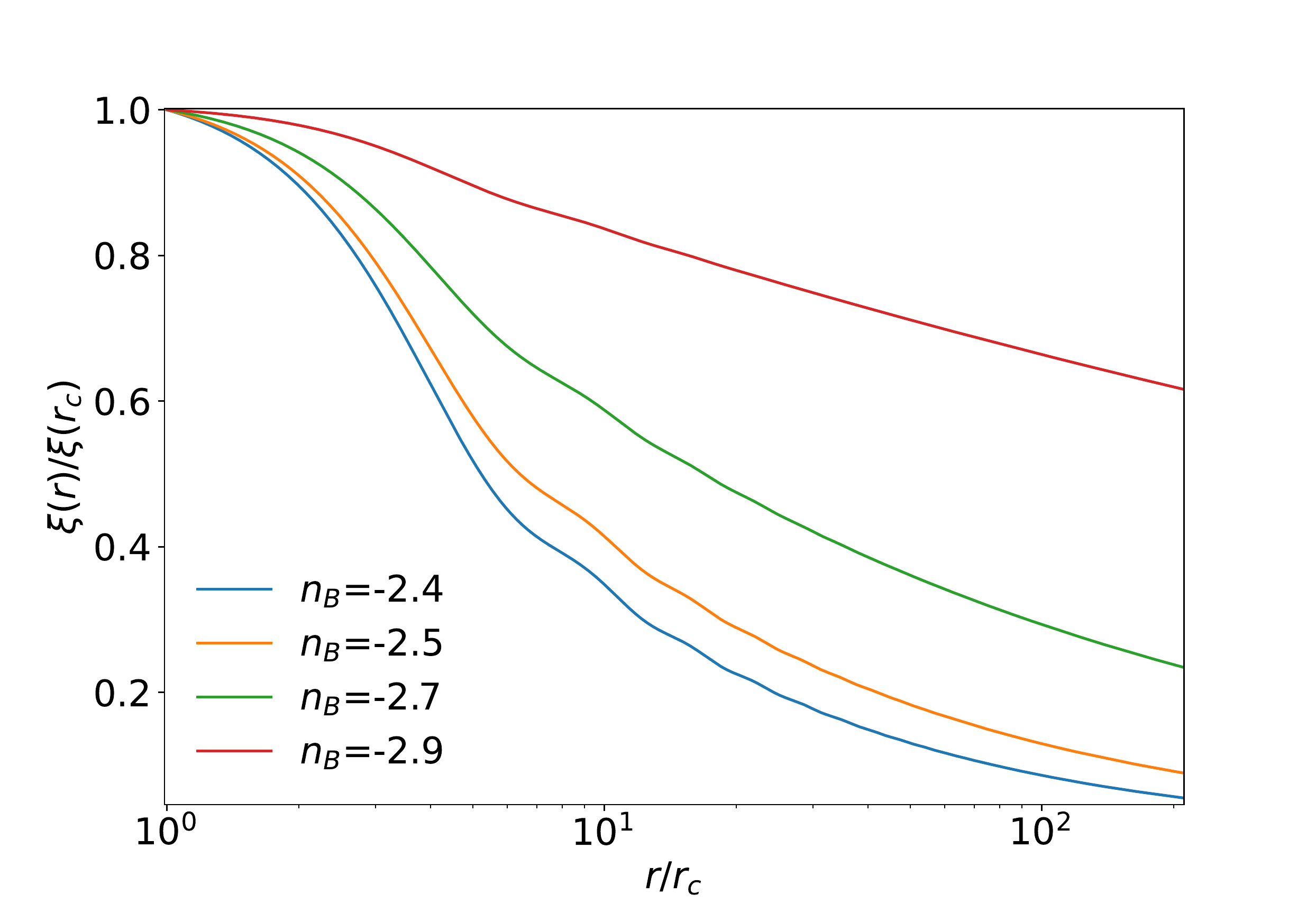}
    \caption{Magnetic field correlations for different values of the spectral index, $n_{_b}$, in real space. The y-axis in figure is $\xi(r)/\xi(r_c)$, and the x-axis is distance in $r_c$ units.}
    \label{fig:xi}
\end{figure}

It is reasonable to assume a power law form of the spectral function, that is,
\begin{equation}
\label{eq:spectrum}
    M(k) = A k^{n_{_b}}
\end{equation}
On the cluster scale, simulations suggest that the spectral index may be roughly equal to $-2.7$ \citep{Dolag:2002}. The parameter $A = V \pi^2 B_0^2 \frac{3+n_{_b}}{k_c^{3+n_{_b}}}$, is fixed by demanding $\sum_{i} \langle B_i(\vec{r})B_i(\vec{r})\rangle = B_0^2$, where $B_0$ is the magnetic field averaged over the scale $r_c$. We add that in our model we assume one-to-one mapping between radio galaxy orientation and the magnetic field and thus the magnitude  $B_0$ remains insignificant in our analysis. Relaxing this assumption or considering $B_0$ dependent mapping between radio galaxy orientation and the magnetic field or both will result in a weaker alignment.
Choosing a coordinate system such as $\vec k$ points in $\hat z$ direction,  $P_{xx}=P_{yy}=1$ and all other components $P_{ij}=0$. The only non-zero terms, $xx$ and $yy$ are therefore, 
\begin{eqnarray}
\label{eq:xi}
\xi_{xx}(r)&=&\langle B_x(\vec r_0+\vec r) B_x(\vec r_0)\rangle, \nonumber \\ 
            &=&\frac{1}{V} \int_{0}^{k_c} A k^{n_{_b}} \left(\int_0^{2\pi} d\phi \int_{-1}^{1} e^{i k r \cos (\theta)} d \cos(\theta) \right)k^2 dk \nonumber \\
            &=& \frac{2\pi A}{V} \int_{0}^{k_c}  k^{2+n_{_b}} \left( \int_{-1}^{1} e^{i k r \cos (\theta)} d \cos(\theta) \right)dk, \nonumber \\
            &=& \frac{4\pi A}{V} \int_{0}^{k_c}  k^{2+n_{_b}} \left( \frac{\sin(kr)}{kr} \right)dk 
\end{eqnarray}
We substitute $A = V \pi^2 B_0^2 \frac{3+n_{_b}}{k_c^{3+n_{_b}}}$ in equation \ref{eq:xi}, and plot magnetic field correlations in figure \ref{fig:xi}.

\section{Procedure}
\label{sec:proc}
We generate the correlated magnetic field and assume the magnetic field direction as a proxy to galaxy PA. For generating the magnetic field, we follow the procedure described in \cite{Agarwal:2012} and in \cite{Tiwari:2016sp}. We consider 3D discrete space consisting of a large number of grids of equal size. The magnetic field is assumed to be uniform in each grid. The magnetic field in the k-space is uncorrelated, with its component $b_k$ along $\vec{k}$, equal to zero. For simplicity, we use polar coordinate ($\vec k$, $\theta$, $\phi$), then the remaining two orthogonal component of the magnetic field in k-space, $b_\theta$ and $b_\phi$ can be considered independent. Next, the k-space magnetic field is obtained  by setting $b_k=0$, and generating $b_\theta$, $b_\phi$ by assuming  normal distribution:
\begin{equation}
    \label{eq:k-field}
    f(b_\theta(k),b_\phi(k)) = N \exp\left[- \frac{b_\theta ^2(k) + b^2_\phi (k)}{2 M(k)} \right], 
\end{equation}
where, N is the normalization factor. We next obtain the real space magnetic field by Fourier transforming the k-space field. We consider a large volume in our simulations, and thus $r_{\rm max} \approx \infty$, i.e., the lower limit of the integration in equation \ref{eq:bcorr_real}, $k_{\rm min}=r^{-1}_{\rm max}$ is set to zero. The real space magnetic field thus obtained follows the correlations governed by field spectrum in equation \ref{eq:spectrum}. 

\subsection{Correlation of galaxies inside clusters}
\label{ssec:proc_cluster}
\begin{figure}
    \centering
    \includegraphics[width=0.7\columnwidth]{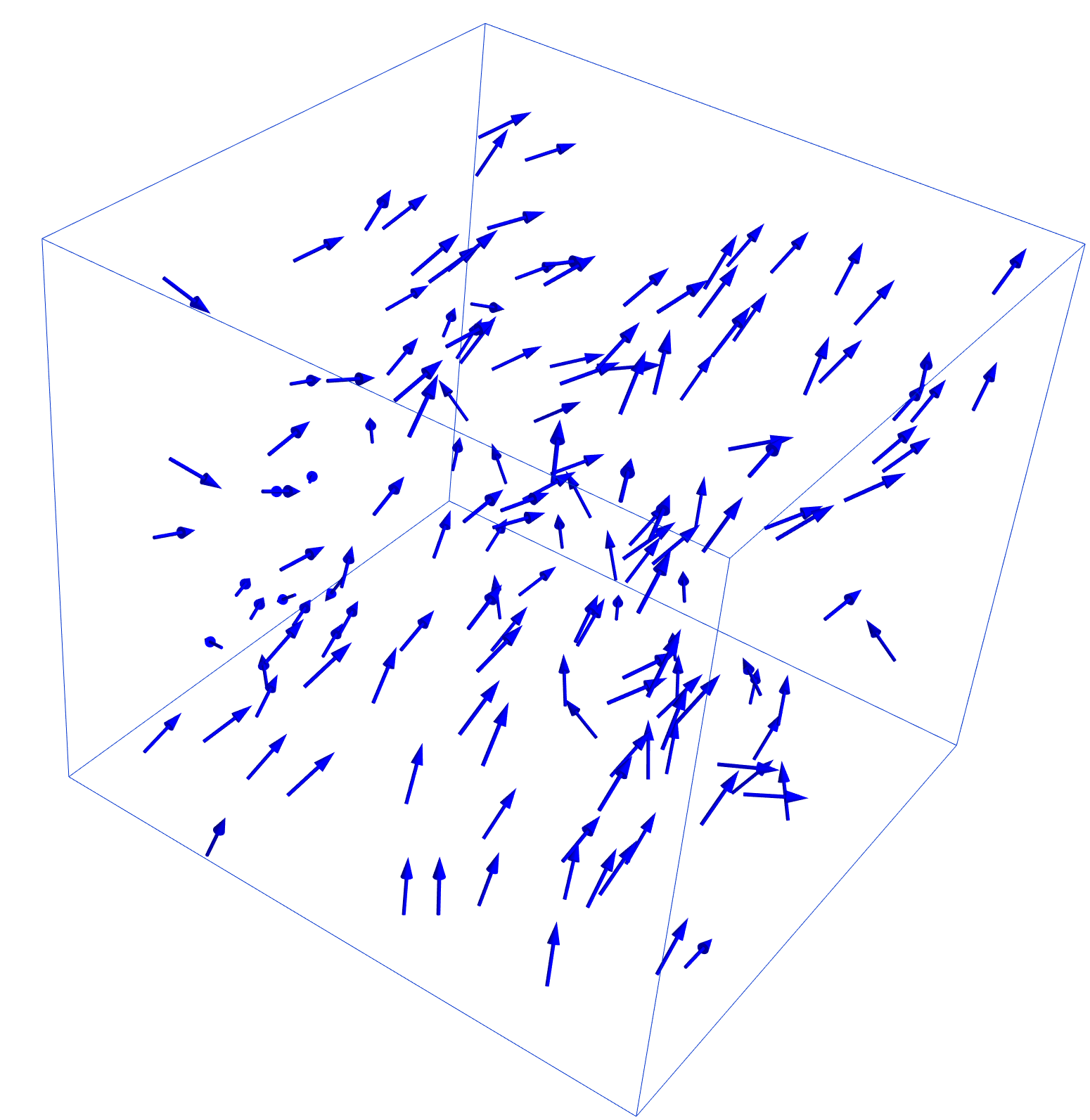}
    \caption{Simulated magnetic field angles inside a 3 Mpc box. }
    \label{fig:3mpcbox}
\end{figure}

\cite{Tovmassian:2020} studied the galaxies from the ACO \citep{ACO:1989} clusters. They consider galaxies within 1 Mpc from cluster center and galaxies in the ring with radii 1 to 2 Mpc. To emulate the galaxy PAs inside the clusters we generate correlated magnetic field over a $1024$ cube grid size box. Each grid size is 0.1 Mpc and in total we simulate the magnetic field over a volume approximately equal to  $1.07 \times 10^6$ $\mpcc$. We randomly choose 100 positions (clusters)  from our simulated box. Next we look around 1 Mpc and 1-2 Mpc radii around these locations and randomly select 100 positions (galaxies), the magnetic field vector at these positions serves as a proxy for the galaxy PAs. A 3 Mpc cubic box from our simulation is shown in figure \ref{fig:3mpcbox}, and the magnetic field vectors on grids, inside this box, are drawn to give  realization of magnetic field alignment obtained in our simulation. Note that in our simulation we have a resolution of {$r_c=0.1$} Mpc and within 1 Mpc radius sphere there are $\sim$ 4188 mini cubes each of side 0.1 Mpc and we randomly chose 100 out of these; the magnetic field in these boxes represent the PA of a galaxy. 

In total we generate 10 full simulations for each spectral index, and from each simulation we randomly find 100 clusters, so in total we simulate the galaxy PAs for 1000 cluster positions (for each spectral index), each containing 100 galaxies and the magnetic field indicating its PA. The 98 clusters  explored by  \cite{Tovmassian:2020} contain roughly 40-60 galaxies within 2 Mpc radius. We exactly mimic the galaxy counts in these clusters in our simulation. That is to mock a  given cluster with $n$ galaxies, we take a mock cluster and chose $n$ galaxies (out of 100) randomly. Next we follow alignment definition\footnote{i.e.  alignment is assumed if in a cluster the galaxy number at one $90^\circ$ position angle interval is more than twice higher than the other $90^\circ$ interval} given in \cite{Tovmassian:2020} and compute alignment. We  perform our analysis for spectral index $n_{_b}=-2.3$ to $-2.98$. 

\subsection{Analysis of the FIRST catalog galaxies}
\label{ssec:FirstAnalysis}

Following \cite{Panwar:2020} and  earlier studies \citep{Agarwal:2012,Tiwari:2013pol,Tiwari:2018} we consider the alignment statistics as follows. Consider a source situated at site $k$ (in real space) with its polarization angle $\psi_k$. Let there be $n_v$ sources within an angular separation $\Delta \theta$ from this source. Next let $\psi_i$  be the polarization angle of 
the source at $i^{\rm {th}}$ site within the $n_v$ sources. A dispersion measure 
of polarization angles relative to $k^{\rm {th}}$ source  $\psi_k$ with its $n_v$  
neighbours is written as, 
\bea
d_{k} = \frac{1}{n_{v}} \sum_{i=1}^{n_{v}} \cos[2(\psi_{i}+\Delta_{i\rightarrow
k}) - 2{\psi}_{k}],
\label{eq:dispersion}
\eea
where $\Delta_{i\rightarrow k}$ is a correction to angle $\psi_{i}$ due to its parallel transport from site $i\rightarrow k$ \citep{Jain:2003sg}. This correction is significant at large scales and cannot be ignored. The position angles span a range of $0$ to $180^\circ$ and to make them behave like usual angles i.e. values in range $0$ to $360^\circ$ we multiply the polarization angles by two \citep{Ralston:1999}. 
The function $d_k$ in above equation \ref{eq:dispersion} is the average of the cosine of the differences of the polarization vector at site $k$ and those of its $n_v$ neighbours within an angular separation $\Delta \theta$ from site $k$. It provides a measure of the dispersion in angles. It is higher for data with lower dispersion and vice versa. 
We take the average of $d_k$ over all source sites and define this 
as a measure of alignment in sample over an angular scale of $\Delta \theta$, 
\bea
S_{D} = \frac{1}{N_t} \sum_{k=1}^{N_{t}} d_{k},
\label{eq:statistics}
\eea
where $N_t$ is the total number of sources in the sample.  Similar alternate statistics 
can also be defined as a measure of alignment \citep{Bietenholz:1986,Hutsemekers:1998,Jain:2003sg,Tiwari:2013pol,Pelgrims:2014}. 
Note that the statistics $S_D$ is non zero even for random PAs, nevertheless the significance of alignment can be obtained comparing the data $S_D$ with that obtained from random PA distributions. We recover the alignment statistics seen in \cite{Panwar:2020} and obtain significant alignment (more than approximately 2 sigma) up to an angular scale of $1^\circ$.

To analyse FIRST correlations, we consider the data prepared in \cite{Panwar:2020}. After considering all the  masks,  cuts from \cite{Panwar:2020}
and an integrated flux cut of 3 mJy, we obtain 5619 sources in total. 
 \cite{Panwar:2020} also performed a three dimensional analysis. For this purpose they used the Unified catalog compiled in \cite{Kimball:2008} in order to obtain redshift information of radio sources. The Unified catalog contains radio sources from NRAO VLA1 Sky Survey (NVSS; \citealt{Condon:1998}), Westerbork Northern Sky Survey (WENSS; \citealt{Rengelink:1997}), Green Bank 6 cm survey (GB6; \citealt{Gregory:1996}) and Sloan Digital Sky Survey (SDSS; \citealt{York:2000,Stoughton:2002,Adelman-McCarthy:2007},  and references therein)  besides the FIRST and hence has more sources in comparison to the FIRST catalog. A total of 593 sources were identified for which redshift information is available. We include all of these sources in our analysis in order to maximize the number for sources for which redshift information is available. This leads to a total of 5941 sources which is slightly larger than those considered in \cite{Panwar:2020}. We have verified that this small change does not produce any significant change in the statistic.

\begin{figure}
    \centering
    \includegraphics[width=\columnwidth]{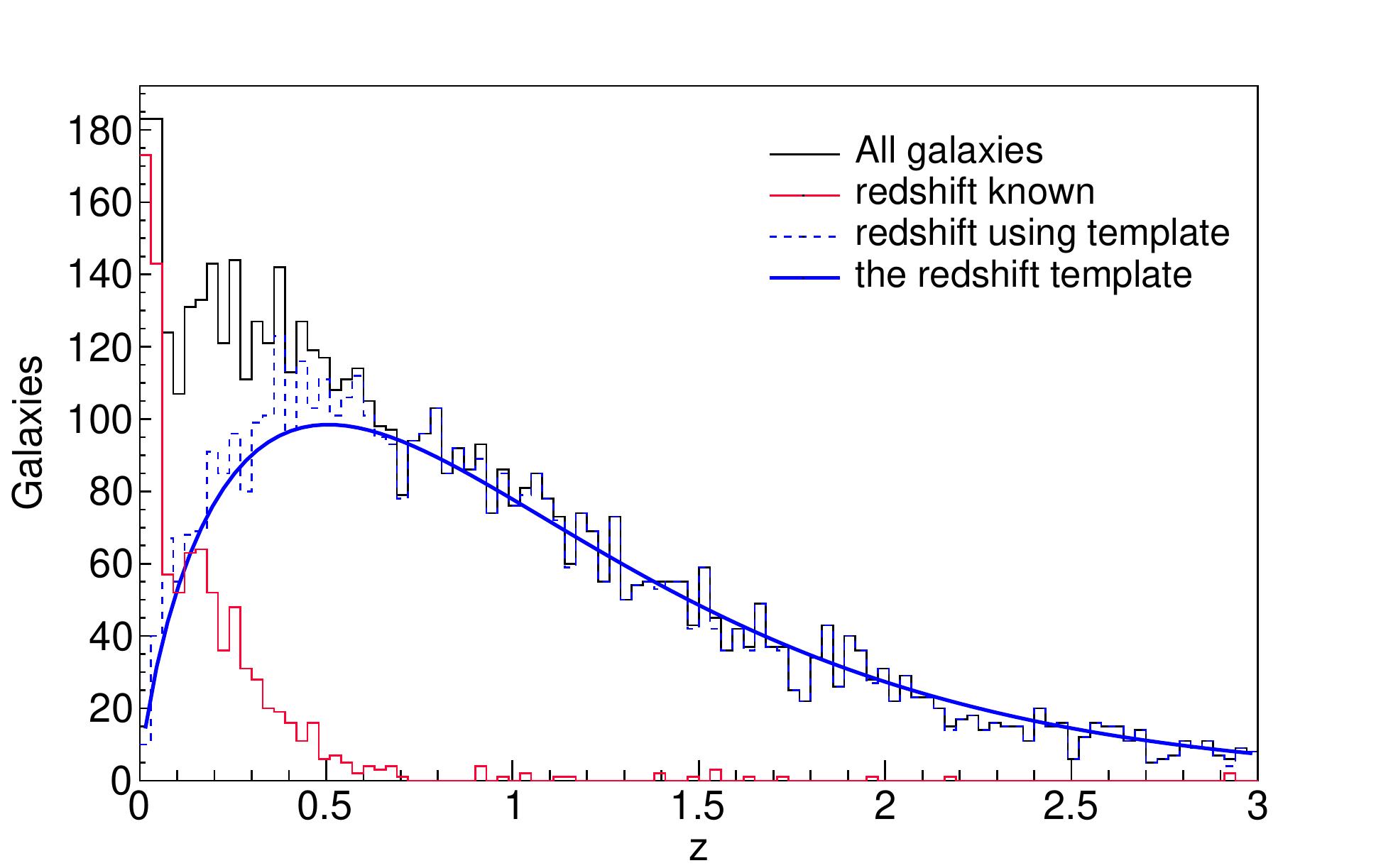}
    \caption{The redshift distribution of FIRST galaxies in our simulation. The red histogram shows the galaxy count with known redshift. For each galaxy with unknown redshift, we assume the blue curve \citep{Adi:2015nb} probability distribution and randomly draw its redshift from this function. The resultant redshifts for one realization are shown as blue dashed histogram. The black histogram shows redshifts thus prepared for all 5941 sources.}   \label{fig:FIRST_z}
\end{figure}

For emulating PA's of FIRST catalog galaxies,  we follow the analysis in \cite{Tiwari:2016sp} and simulate magnetic field in 3D space for spectral index values in the range $-2.30$ to $-2.98$. The median redshift of FIRST galaxies is about 0.8 \citep{Condon:1989} and the redshift information for 593 sources out of 5941 is available 
\citep{Kimball:2008}. The FIRST galaxies spans a huge volume in space and to incorporate the sample we simulate the magnetic field up to redshift 3, i.e., a comoving radial distance of  approximately 6.5 Gpc. Note that with our $1024$ cube grid size simulation, this huge volume is obtained by scaling the grid size appropriately. In particular to generate magnetic field simulation up to redshift 3, we set our grid size to be approximately $r_c=13$ Mpc. For each spectral index value we generate 100 magnetic field simulations. For our analysis we need to glean out the magnetic field information for each 5941 galaxy position in 3D space and therefore need the radial (redshift) distance information along with angular positions. As mentioned above, we only have redshifts for 593 sources out of 5941, for the remaining 5348 we rely on  the radio source redshift profile prepared using the Combined EIS-NVSS Survey of Radio Sources (CENSORS, \citealt{Best:2003,Rigby:2011})  and the Hercules \citep{Waddington:2000,Waddington:2001} data sets. We argue that this is the best redshift template (see figure \ref{fig:FIRST_z}) we expect for radio galaxies, the same has been used in several other studies \citep{Rigby:2011,Adi:2015nb,Tiwari:2016adi,Tiwari:2021}.

\begin{figure}
    \centering
    \includegraphics[width=\columnwidth]{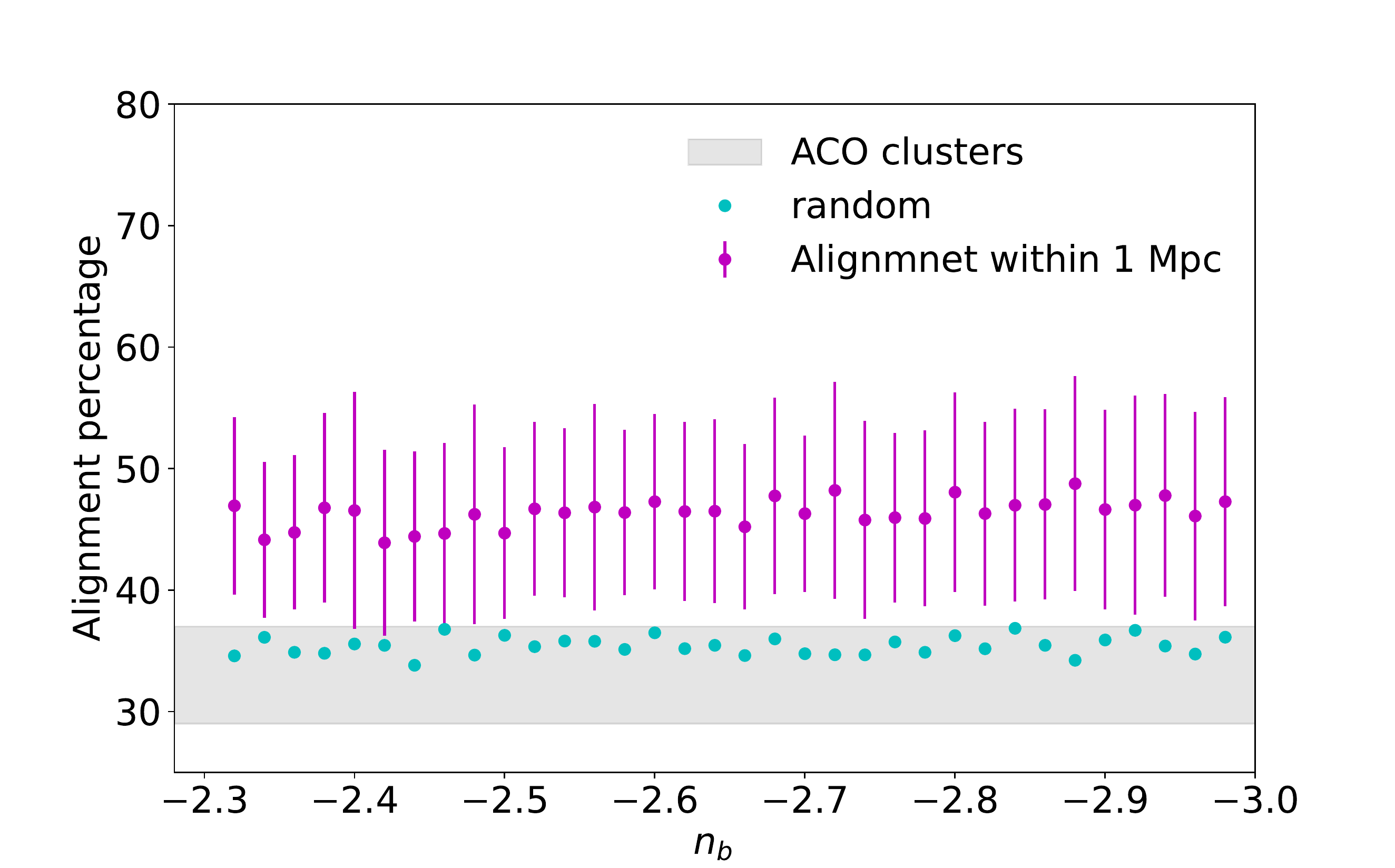}
    \caption{Alignment versus spectral index observed with simulation inside cluster within 1 Mpc radius. The random (cyan) also have similar error bars as the data mock (magenta), though we don't plot to avoid overfilling the figure. The gray band shows the observations by \protect\cite{Tovmassian:2020}.}%
    \label{fig:1Mpc_align}
\end{figure}
\begin{figure}
    \centering
    \includegraphics[width=\columnwidth]{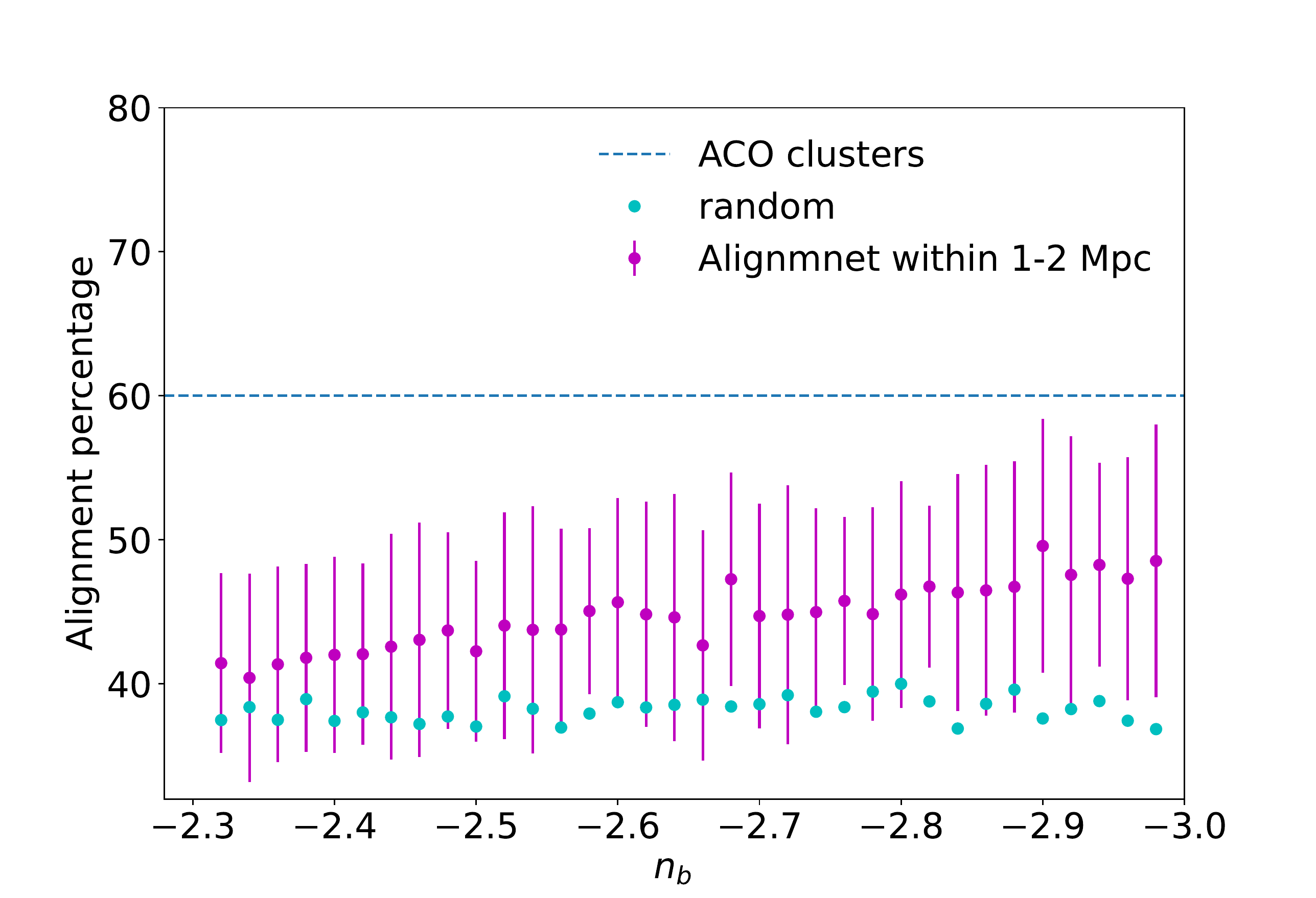}
    \caption{Alignment versus spectral index observed with simulation in cluster radii 1-2 Mpc range. The dashed line shows the observations by \protect\cite{Tovmassian:2020}. The mock (magenta) and random (cyan) detail are same as in figure \ref{fig:1Mpc_align}.  }
    \label{fig:2Mpc_align}
\end{figure}

\section{Results}
\label{sec:results}
\subsection{Alignment inside cluster }
\label{ssec:res_cluster}
We follow the procedure described in Section \ref{ssec:FirstAnalysis} and observe that the alignment signal doesn't change much with spectral index for galaxies within 1 Mpc radius. However for radii 1-2 Mpc the alignment signal increases  with spectral index. \cite{Tovmassian:2020} study a limited number of clusters, e.g. they consider 73 clusters, and find alignment in 43 clusters for radii 1-2 Mpc, with our simulation we consider sets of exactly same numbers of clusters and find percentage clusters showing alignment.  Next, using the results from these sets we get the mean and standard deviation (error bars) on alignment percentage. We observe approximately 45\% alignment within 1 Mpc radius of clusters and up to 48\% alignment for galaxies within radii 1-2 Mpc.  \cite{Tovmassian:2020} report about 60\% alignment for galaxies in  cluster radii 1-2 Mpc range, and about 37\% to 29\% alignment (depending on cluster richness) for galaxies within 1 Mpc cluster radius. The results are shown in figures \ref{fig:1Mpc_align} and \ref{fig:2Mpc_align}.   Although with our model we do observe alignment signal for galaxies inside cluster, we mismatch the right amplitude of alignment. Namely, for galaxies within 1 Mpc inside cluster the alignment signal with our model is somewhat higher; and on the other hand even with most negative $n_{_b}$ we don't observe the alignment as high as seen by \cite{Tovmassian:2020}. We argue that observed alignment signal could be different in magnitude, because the galaxy evolution and  gravitational interactions between galaxies can dramatically change alignment at different scale. Even so we do observer similar alignment signal with our model.

\subsection{FIRST catalog galaxies}
\label{ssec:res_first}
\begin{figure}
    \centering
    \includegraphics[width=\columnwidth]{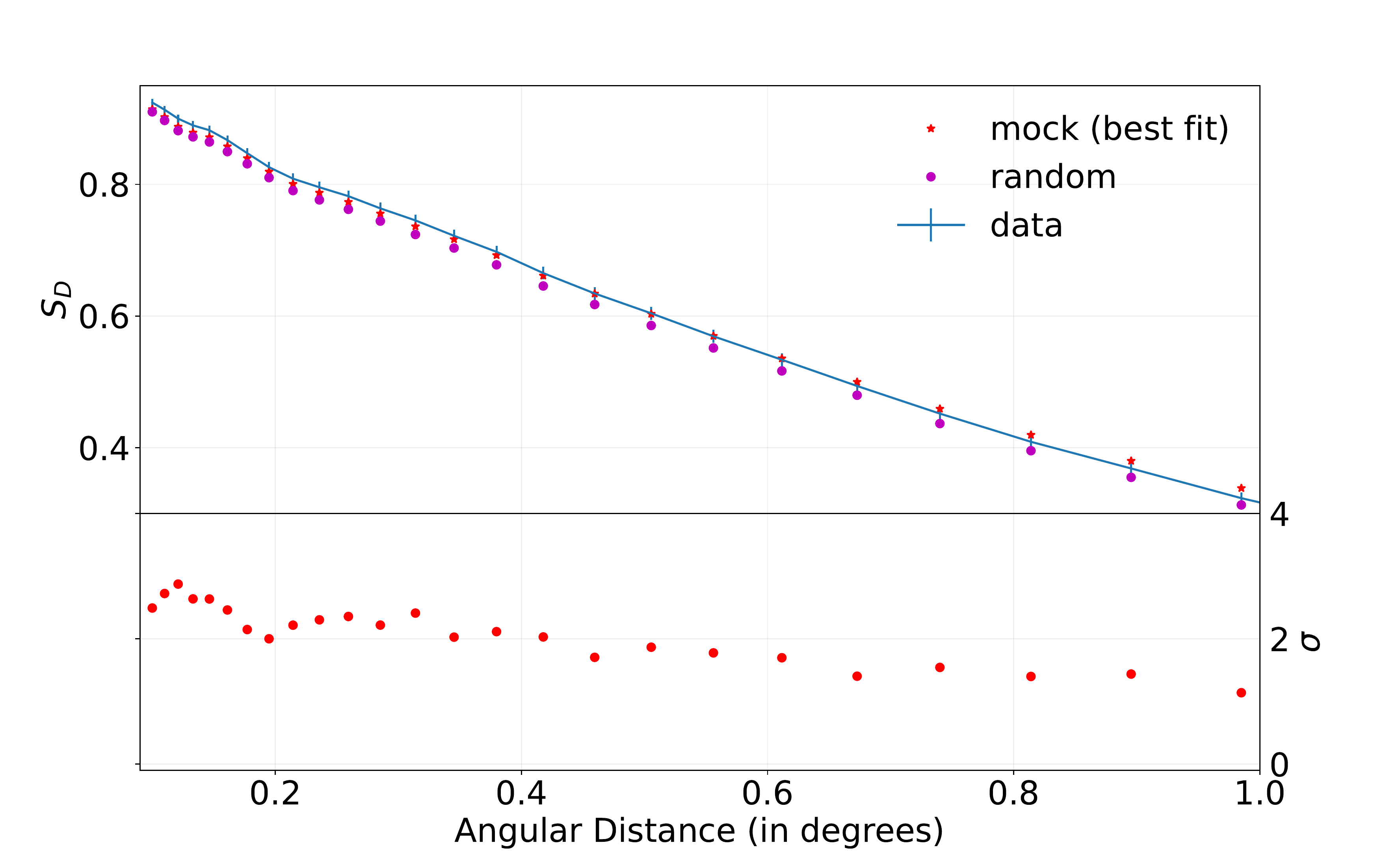}
    \caption{The cyan color shows the data $S_D$ and the red asterisk shows the $S_D$ for mock catalog. The mock catalog generated assuming magnetic field correlations with spectral index $n_{_b}=-2.62$ and best fits with data. The magenta circles are the $S_D$ observed with random PAs, the difference between magenta circles and red data points is the alignment signal.}
    \label{fig:FIRST_fit_1deg}
\end{figure}

\begin{figure}
    \centering
    \includegraphics[width=\columnwidth]{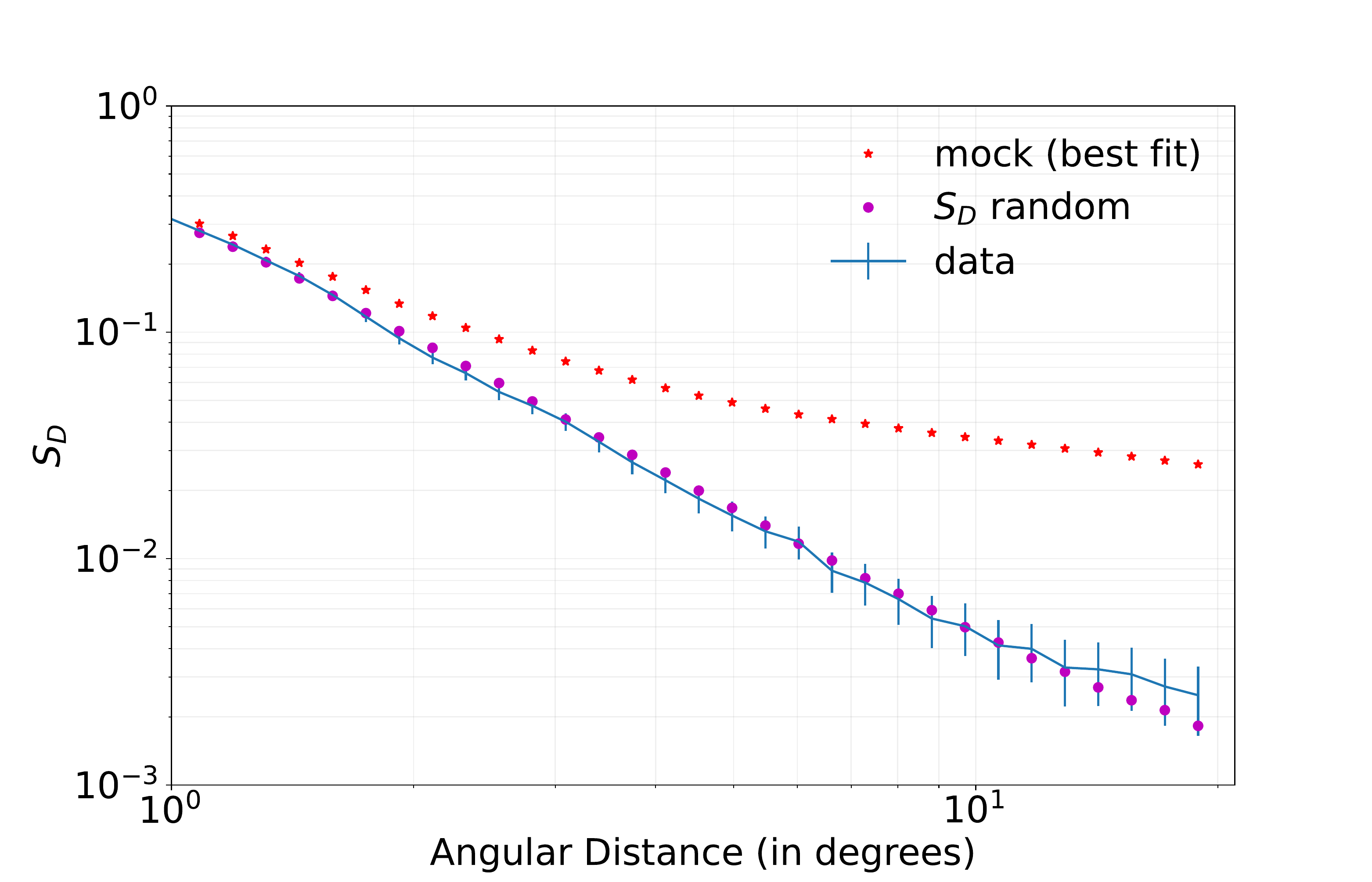} 
    \caption{Alignment at relatively large scales. Note that the magnetic field (and thus the PAs from mock) shows relatively significant correlation at higher scales. }
    \label{fig:FIRST_fit}
\end{figure}

We attempt to explain the the alignment signal observed in \cite{Panwar:2020} and calculate the statistics $S_D$ for emulated catalogs and PAs obtained using magnetic field simulation discussed in section \ref{ssec:FirstAnalysis}. For a range of spectral index $n_{_b}$ we produce 100 catalogs and corresponding PAs and obtain theoretical $S_D$ up to $1^\circ$ angular scale where we expect to see the alignment. We estimate the jackknife error in the  statistic $S_D$ as we do not know the exact errors in polarization measurements. We compare the data with random and obtain $2-3\sigma$ alignment signal up to $1^\circ$ angular scale. The results are in figure \ref{fig:FIRST_fit_1deg}, albeit the information about small scale alignment is not easily 
perceptible. This is because the statistic $S_D$ is calculated as a function of distance between sources. As this  distance becomes small, the number of nearest neighbours of any source decrease. Hence for sufficiently small distances, the signal of alignment is diluted since $S_D$
dominantly contains the correlations of each source with itself. This is why the random and correlated samples appear to be close to one another at small angular distances. Even so, we point out that for the lower cutoff used here, the data still shows a significant signal.
The best value of $n_{_b}$ is obtained by making a $\chi^2$ fit with the observed data $S_D$ and mock $S_D$ for a range of angular distances. We define $\chi^2$ as,
\begin{equation}
\label{eq:chi2}
\chi^2= \sum_i \sum_j (S_D^{\rm data}-S_D^{\rm mock})_i (\mathbf{C}^{-1})_{ij} (S_D^{\rm data}-S_D^{\rm mock})_j 
\end{equation} 
where the subscripts $i,j$ represent the quantities over angular distances $\theta_i$ and $\theta_j$, respectively. $\mathbf{C}$ is the covariance matrix determining the uncertainties and correlations in measured $S_D$ at different angular distances. To obtain $\mathbf{C}$ we generate 1000 random data sets by shuffling the PAs among different sources\footnote{Alternatively, one can generate random PAs in 0 to $\pi$ range. However, by shuffling the PAs among different sources we ensure that every random set follows the same PA distribution.}. Next, we obtain $S_D$ for each random data set and calculate, 
\begin{equation}
  \mathbf{C}_{ij}= \langle (S_D)_i (S_D)_j \rangle -\langle(S_D)_i\rangle \langle(S_D)_j\rangle  
\end{equation}
We notice that the diagonal element of the covariance matrix $\mathbf{C}$ from random sets, i.e., the variances of $S_D$s, are significantly smaller in comparison with jackknife variance estimates. The covariance matrix from random data sets count for ``standard error" that is the standard deviation of sampled data (available data) statistics. In contrast, the jackknife error estimates presumably count for systematics in addition and thus present more complete value for uncertainties present in $S_D$ determination. We multiply the covariance matrix appropriately so that its trace, i.e., sum of its diagonal elements, matches the jackknife variances sum for our fit range. We fit i.e. sum up to $1^\circ$ as the alignment signal is observed only up to this scale. We find that $n_{_b}=-2.62\pm0.03$ best explains the observed alignment with $\chi^2$/dof equal to 1.35. If we ignore the correlation  of $S_D$s and consider the jackknife variances, i.e., $\mathbf{C_{ij, i\ne j}} =0$ and  $\mathbf{C}_{ii}=Var((S_D)_i)$  using jackknife method, we find  $\chi^2$/dof equal to 1.31. We add that we are fitting 25 data points and with one parameter the number of degrees of freedom (dof) is 24. The results are shown in figures \ref{fig:FIRST_fit_1deg} and \ref{fig:FIRST_fit}. The mock catalog generated using correlated magnetic field as a proxy also shows the significant alignment beyond $1^\circ$ scale (figure \ref{fig:FIRST_fit}).  However, the data beyond $1^\circ$ shows no alignment and agrees with random. This may arise due to insufficient redshift information. Due to unavailability of redshifts, we are only able to analyse 2D alignments. As we go to larger angular distances, the sources have higher probability of being at different redshifts, and thus introducing more and more uncertainty in their absolute distances. Consequently, the galaxies at angular scales around 10 degree may have significant fraction of galaxies at relative high and low redshift and the distances between these may be significantly high. For example, galaxies sitting at redshift 0.1 (very low redshift) and redshift 2 (very high redshift!) even with no angular separation are separated by several Gpc and we expect negligible correlation between these. The 2D analysis thus unreliable at large angular scales. There could of course be other reasons too, for example, galaxy interactions and evolution that may destroy alignment at large scales. We do not clearly understand this with present data. We  add that the FIRST data also shows some fluctuations in PAs distribution \citep{Contigiani:2017,Panwar:2020} which we do not understand clearly. Nonetheless, this is taken care while calculating alignment statistics $S_D$ by matching PA distribution seen in data to randoms.

\section{Discussion}
\label{sec:dis}
Our results show that we are able to explain the significant signal observed in the FIRST catalog for angles less than 1 degrees using with $n_{_b}=-2.62$. With this value we are also able to explain the alignment seen inside clusters over distance scales less than 2 Mpc to some extent. However the alignment signal in this case shows a relatively mild dependence on the spectral index and does not necessarily require the value $n_{_b}=-2.62$. Furthermore, the same model explains the alignment signal observed in JVAS/CLASS data \citep{Tiwari:2013pol,Tiwari:2016sp} with slightly higher spectral index equal to $-2.74\pm0.04$. Apparently it seems that  the galaxy alignments at all scales and at all distances (redshifts) can be modeled as a power law in k-space. The best fit spectral index slightly differs for different data sets. However, the data is likely to have considerable errors and hence the difference is not too significant. Also, relaxing one-to-one mapping between radio galaxy orientation and magnetic field, and counting on evolution due to structure formation may result a consistent background primordial spectral index $n_b$. Alternatively, it is possible that the power law model is too simple. A simple generalization would be to introduce a scale and position dependent spectral index. One can further generalize the model and consider background magnetic field magnitude, $B_0$, dependent mapping between radio galaxy orientation and magnetic field. With these generalizations the effective value of $B_0$ and $n_{_b}$ could differ for different data sets. 

For the FIRST data, the theoretical estimate of the statistic $S_D$ starts to deviate from data for angles larger than 1 degree. This may arise if the spectral index of the magnetic field changes significantly beyond this angle. Another possible explanation is that here we are using only two dimensional analysis. Hence beyond a certain angular distance it may no longer be reasonable to assume that most the sources within this angular separation are spatially close to one another. A proper analysis is only possible once we have redshift information for these sources. The data set for which this parameter is available is rather limited and hence a detailed analysis is best postponed to future work.

\section{Conclusions}
\label{sec:con}
In this paper we have argued that the alignment seen in the radio galaxy axis over distance scales of few Mpc to several tens of Mpc may be explained in terms of correlations in magnetic field in real space. We emphasize that the magnetic field correlations of similar nature (i.e. approximately same spectral index) can explain  the radio polarization alignment observed with JVAS/CLASS data over 100 Mpc scale \citep{Tiwari:2016sp} and potentially also explain the quasar alignment observed over Gpc \citep{Agarwal:2012}. Hence we find that, remarkably, the magnetic field correlations seem to explain observed alignment signals over all scales. The magnetic field is assumed to be uncorrelated in Fourier space, i.e. it is proportional to $\delta_{\vec k,\vec q}$, with its two point correlations described by a spectral function $M(k) = Ak^{n_{_b}}$ with the spectral index $n_{_b}<0$. Theoretical simulations suggest $n_{_b}\approx -2.7$. The magnetic field acquires correlations over large distances in real space, which increase with increase in the magnitude of $n_{_b}$, i.e. as the spectral index becomes more negative. We assume a simple model in which we set the galaxy axis along the direction of the background magnetic field. A future refinement of our work would be to relate the alignment in galaxy axis to those in the magnetic field through simulations. This will allow us to go beyond our assumption that the galaxy axis points along the background magnetic field direction. 
Using the model described above, we determine the spectral index that fits the data for the alignment signal seen in galaxies inside clusters on distance scale of 2 Mpc \cite{Tovmassian:2020} as well as on large scale of about 28 Mpc, seen in the FIRST catalog \cite{Panwar:2020}. We find that FIRST alignment signal can be explained if we choose the spectral index $n_{_b}=-2.62$. In our model the  alignment within clusters at distance less than 2 Mpc does not show a strong dependence on $n_{_b}$ nonetheless, considering the best fit spectral index using the FIRST or JVAS/CLASS data the model also provides a reasonable explanation of the galaxy alignment at such short distance scales.  The value $n_{_b}=-2.62$ is dominantly selected by analysis of the FIRST data. It is encouraging that the value obtained by this analysis is close to that expected from simulations.
\section{Acknowledgments}
We thank Mohit Panwar for making FIRST catalog used in their work available to us. 
This work is supported by the RFIS grant (No. 12150410322) by the National Natural Science Foundation of China (NSFC) and the support by the National Key Basic Research and Development Program of China (No. 2018YFA0404503) and NSFC Grants 11925303 and 11720101004 and through the DST-SERB grant number EMR/2016/004070. 
\section{Data availability}
The data that support the analysis and plots within this paper are available from the corresponding author upon reasonable request.

\bibliographystyle{mn2e}
\bibliography{master}
\end{document}